\def\mOne{0.776}
\def\mTwo{1.230}
\def\mThree{1.465}
\def\integralToEndpt{0.35}
\def\diffWithDRBelowEndpt{0.13}
\def\BornDifference{0.06}
\def\withBorn{0.19}

\def\newIOnefree{0.094}
\def\newIOneFZero{0.103}
\def\RCfree{0.03917}
\def\RCFZero{2.43}
\def\Delfree{0.02396}
\def\DelFZero{2.45}
\def\unitarityMis{2.3}
\def\FQZeroCNTR{0.575}
\def\QZeroCNTR{1.10}
\def\CoefITwoCNTR{1.965}
\def\ConditOne{7.86}
\def\IBorn{0.85}
\def\FZeroCNTR{1.42}
\def\RCBj{0.03942}
\def\DelBj{0.02421}
\def\RCInterpol{0.03952}
\def\DelInterpol{0.02431}
\def\RCAv{0.03947}
\def\DelAv{0.02426}
\def\DelAvPerc{2.426}
\def\OnePlusRCAv{1.03947}
\def\RCFirstAp{0.03895}
\def\DelFirstAp{0.02374}
\def\RCQZero{0.03939}
\def\DelQZero{0.02418}
\def\discrepancy{1.1}
\def\absDiscrepancy{4}
\def\newANP{2.44}
\def\ALead{-1.511}
\def\AF{1.422}
\def\BLead{6.951}
\def\BF{-3.533}
\def\CLead{-4.476}
\def\CF{2.092}
\def\VudBj{0.97391}
\def\VudAv{0.97389}

\def\FQZeroSQRD{\FQZeroCNTR(50)} 
\def\QZeroSQRD{\QZeroCNTR(10)}   
\def\CoefITwo{\CoefITwoCNTR(21)}
\def\FZero{\FZeroCNTR(15)}

\def\lineAv{5}
\def\bound{0.16}
\documentclass[letterpaper,english,prd,preprintnumbers,nofootinbib,showpacs,shortbibliography ,notitlepage]{revtex4-1}
\usepackage[T1]{fontenc}
\usepackage[latin9]{inputenc}
\usepackage{amsmath}
\usepackage{amssymb}
\makeatletter

\usepackage{graphicx}
\usepackage{babel}
\makeatother

\usepackage{geometry}
\begin{document}
\def\delVR{\Delta^V_R}
\def\MS06{\cite{Marciano:2005ec}}
\def\DR{\cite{Seng:2018yzq}}
\def\perk{\cite{Markisch:2018ndu,Dubbers:2018kgh}}
\def\CMSO4{\cite{Czarnecki:2004cw}}
\def\CMS18{\cite{Czarnecki:2018okw}}
\def\aMS{ \alpha_{\overline{\scriptscriptstyle MS}} }
\global\long\def\edge#1{\left.#1\right|}
\global\long\def\d{\mathrm{d}}
\global\long\def\order#1{\mathcal{O}\left(#1\right)}
\def\Vud{V_\text{ud}}
\def\Vus{V_\text{us}}
\def\Vub{V_\text{ub}}
\def\BR{\text{BR}}
\def\sig{$\sigma$}
\def\second{\text{s}}
\def\gev{\text{ GeV}}

\selectlanguage{english}

\preprint{Alberta Thy 7-19}
\title{Radiative Corrections to Neutron and Nuclear Beta Decays
  Revisited}

\author{Andrzej Czarnecki }
\affiliation{Department of Physics, University of Alberta, Edmonton, 
Alberta, Canada T6G 2E1}
\author{William J.~Marciano}
\affiliation{Department of Physics, Brookhaven National Laboratory,
  Upton, New York 11973, USA}
\author{Alberto Sirlin}
\affiliation{Department of Physics, New York University,\\
    726 Broadway, New York, New York 10003, USA}

\begin{abstract}
  The universal radiative corrections common to neutron and
  super-allowed nuclear beta decays (also known as ``inner''
  corrections) are revisited in light of a recent dispersion relation
  study that found $+2.467(22)\%$, i.e.~about $2.4\sigma$ larger than
  the previous evaluation. For comparison, we consider several
  alternative computational methods. All employ an updated
  perturbative QCD four-loop Bjorken sum rule (BjSR) defined QCD
  coupling supplemented with a nucleon form factor based Born
  amplitude to estimate axial-vector induced hadronic
  contributions. In addition, we now include hadronic contributions
  from low $Q^2$ loop effects based on duality considerations and
  vector meson resonance interpolators.  Our primary result,
  $\DelAvPerc(32)\%$ corresponds to an average of a Light Front Holomorphic
  QCD approach and a three resonance interpolator fit.  It reduces the
  dispersion relation discrepancy to approximately $\discrepancy\sigma$ and
  thereby provides a consistency check.  Consequences of our new
  radiative correction estimate, along with that of the dispersion
  relation result, for CKM unitarity are discussed. The neutron
  lifetime-$g_A$ connection is updated and shown to suggest a shorter
  neutron lifetime $< 879$ s. We also find an improved bound on
  exotic, non-Standard Model, neutron decays or oscillations of the
  type conjectured as solutions to the neutron lifetime problem,
  $\BR(n\to \text{exotics})<\bound\%$.
\end{abstract}
\maketitle  
 
\section{Introduction}
Precision tests of the Standard Model (SM) require accurate
calculations of electroweak radiative corrections (RC)
\cite{Kinoshita:1958ru,Berman:1958ti,berman62,Bjorken:1966jh,Abers:1967zzb,Abers:1968zz,Sirlin:1974ni}. For
example, unitarity of the Cabibbo-Kobayashi-Maskawa (CKM) quark mixing
matrix leads to orthonormal relationships among row and column matrix
elements and provides a means to search for indications of ``New
Physics'' via departures from SM expectations.  However, for those
searches to be credible, strong interaction effects must be reliably
evaluated.

Consider the precise CKM first row unitarity condition
\begin{equation}
|\Vud|^2 + |\Vus|^2 + |\Vub|^2 =1. \label{eq1}
\end{equation}
Employing the PDG 2018 average based on super-allowed $0^+\to 0^+$
nuclear beta decays \cite{Tanabashi:2018oca,Hardy:2016vhg},
\begin{equation}
|\Vud| = 0.97420(10)_{\scriptscriptstyle NP}(18)_{\scriptscriptstyle
  RC}, 
\label{eq2}\end{equation}  
as extracted by Hardy and Towner \cite{Hardy:2016vhg}, using a
universal electroweak radiative correction \MS06 (also known as the
``inner'' correction),
\begin{equation}
 \delVR = 0.02361(38), 
\label{eq3}\end{equation}
along with the $K_{\mu 2}/\pi_{\mu 2}$ and $K_{l3}$ weighted average
\cite{Tanabashi:2018oca} of $|\Vus|=0.2253(7)$ and $|\Vus|=0.2231(8)$
respectively, 
\begin{equation}
|\Vus| = 0.2243(9), 
\label{eq4}\end{equation}
(where the uncertainty has been increased by a scale 
factor $S=1.8$ to account for  $K_{\mu 2}$ and $K_{l3}$  inconsistencies)
and negligible $|\Vub|^2 \sim  O(10^{-5})$ implies
\begin{equation}
  |\Vud|^2 + |\Vus|^2 +|\Vub|^2 =0.9994(4)_\text{ud} (4)_\text{us} ,
\label{eq5}\end{equation} 
Alternatively, one may employ the updated  \cite{Aoki:2019cca}  $K_{\mu 2}/\pi_{\mu 2}$
constraint $|\Vus|/|\Vud|=0.2313(5)$ to derive the unitarity condition \cite{Marciano:2004uf},
\begin{equation}
 |\Vud|=0.97427(11) 
\text{   Unitarity condition from $K_{\mu 2}/\pi_{\mu 2}$.}
\end{equation} 
Both eq.~\eqref{eq2} and eq.~\eqref{eq5}
are in good accord with those Standard Model (SM)
expectations.   However, that
confirmation has recently been questioned. A new analysis of the
universal radiative corrections to neutron and super-allowed nuclear
beta decays based on a dispersion relations (DR) study of hadronic
effects by Seng, Gorchtein, Patel, and Ramsey-Musolf
\cite{Seng:2018yzq} finds a roughly 0.1\% larger
\begin{equation}
  \delVR = 0.02467(22), \label{eq6}\end{equation} 
with reduced uncertainty. It leads to a smaller more precise \DR
\begin{eqnarray}
|\Vud| = 0.97370(14) & & \text{DR result  \protect\DR},
 \label{eq7}\\
|\Vud|^2 + |\Vus|^2 +|\Vub|^2  & = &0.9984(3)(4). 
\label{eq8}\end{eqnarray} 
Both eq.~\eqref{eq7} and eq.~\eqref{eq8}  
exhibit an apparent roughly $3.2\sigma $ violation
of unitarity. Taken literally, it could be interpreted as a strong
hint of ``new physics.'' However, nuclear structure effects and other
corrections to $\Vud$ and $\Vus$ are still being investigated
\cite{Seng:2018qru,Gorchtein:2018fxl}. 

Although the use of DR for such an analysis represents a major
advancement in the calculation of electroweak radiative corrections,
it is important to reexamine the input leading to eq.~\eqref{eq6} and
compare with other computational approaches. In that way, one can
better assess their consistency and individual reliabilities. Close
examination may reveal issues with the RC or other inputs. For that
reason, we update here an alternative study of the radiative
corrections to neutron and super-allowed nuclear beta decays, estimate
hadronic uncertainties and discuss various possible implications.

Before going into detail, let us briefly preview our study. We first
recall the lowest order one loop universal radiative corrections to
neutron and super-allowed nuclear beta decay rates in the framework of
the SM. Leading log QED effects, beyond one loop order, controlled by
the renormalization group are included. Overall, they increase the RC
by about 0.1\%. However, some care must be exercised in examining
compound effects, particularly since the DR result to be compared with
differs from the earlier calculations by a similar $\sim 0.1\%$. That
difference could be offset by smaller changes in several other
contributions to the decay rates.

Consider the weak vector amplitude stemming from tree and loop level
effects. At very low momentum transfer, vector current induced effects
are protected from strong interactions by vector current conservation
(CVC). Hadronic effects, nevertheless, enter the vector amplitude via
$\gamma W$ box diagrams (and to a lesser extent $ZW$ box diagrams),
see Fig.~\ref{fig:box}, where the operator product expansion of quark
axial and vector currents can produce a vector amplitude. In that way,
short-distance QCD and long-distance hadronic structure dependence are
induced by the non-conserved axial current.

\def\skala{0.29}
\begin{figure}[ht]
  \centering
  \begin{tabular}{c@{\hspace*{5mm}}c@{\hspace*{5mm}}c}
  \includegraphics[scale=\skala]{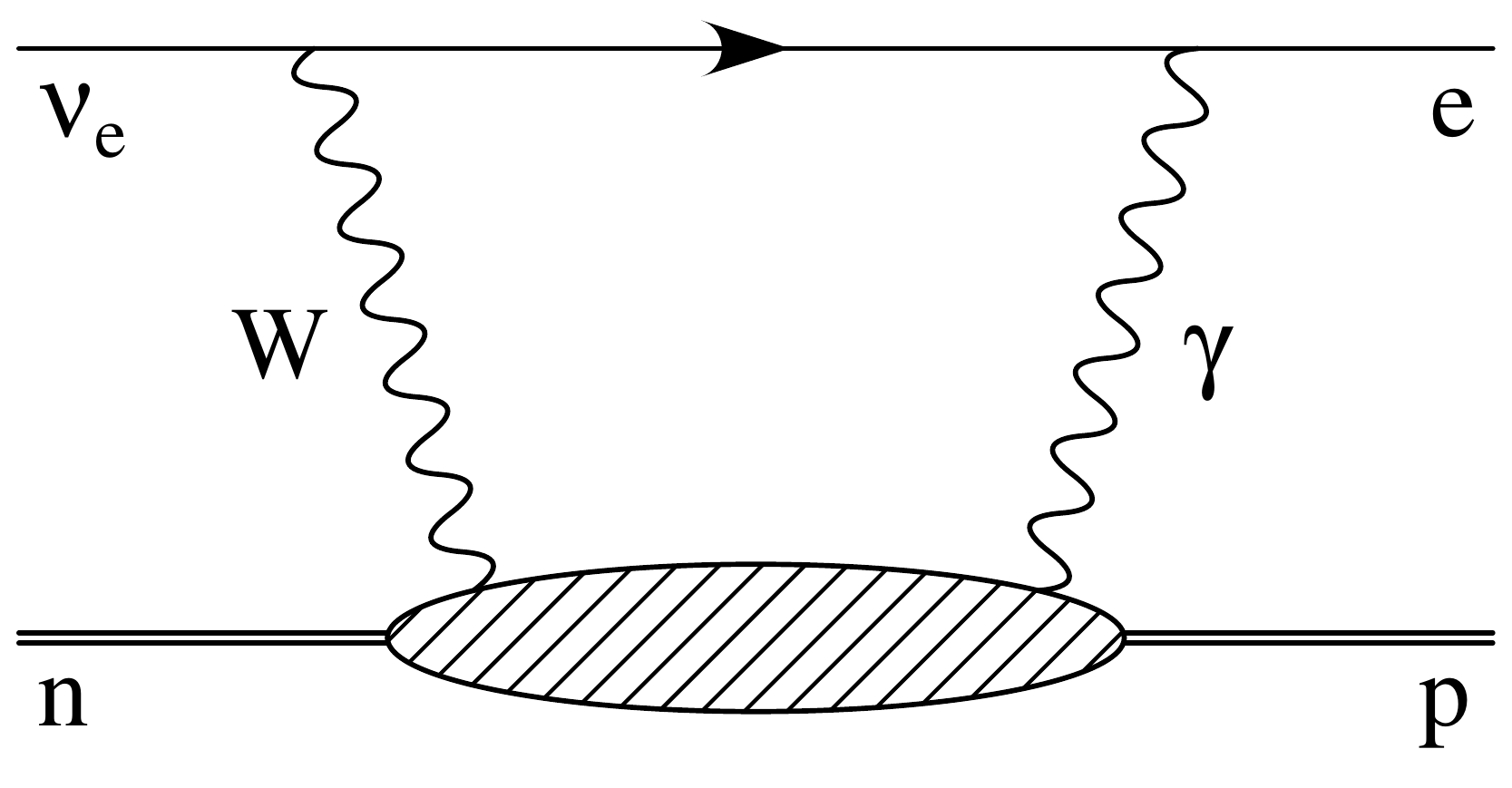} &
  \includegraphics[scale=\skala]{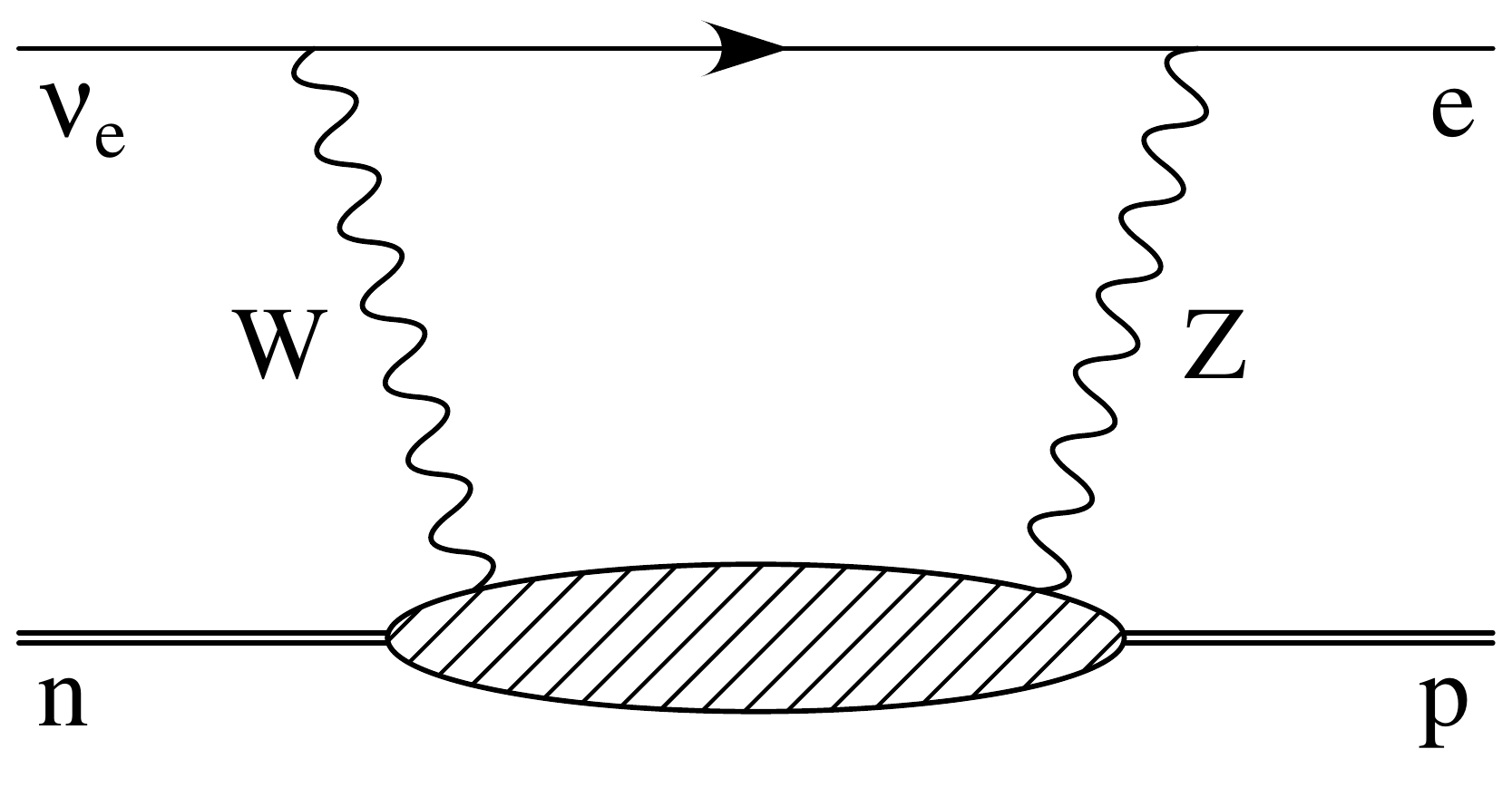} &
  \includegraphics[scale=\skala]{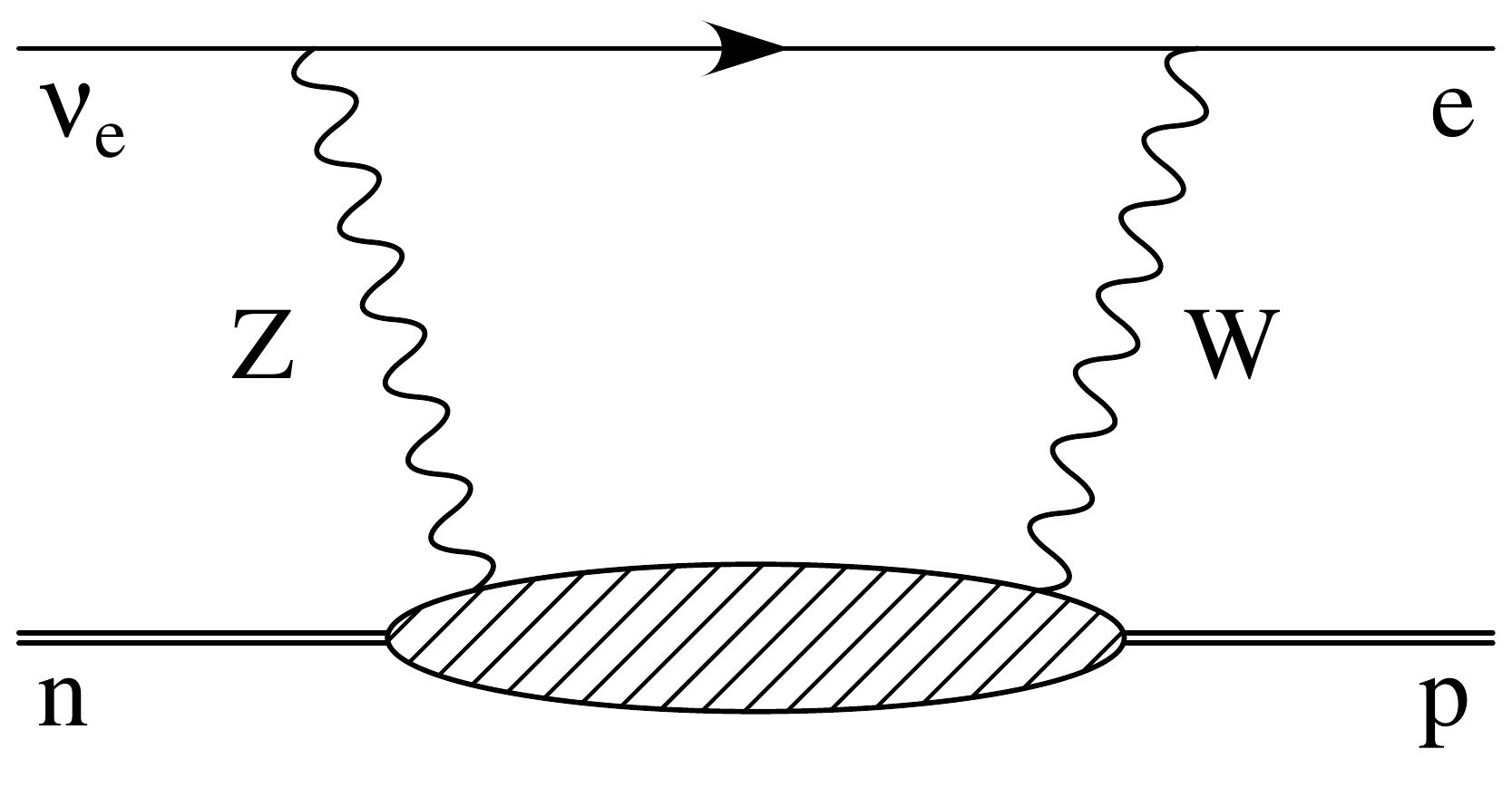}
  \end{tabular}
  \caption{$\gamma W$ and $ZW$ box corrections to  neutron decay.}
  \label{fig:box}
\end{figure}

Up until 2006 \cite{Marciano:2005ec}, only the lowest order,
$\order{\alpha_s}$, QCD perturbative correction to the box diagrams
was considered
\cite{Sirlin:1978sv,Adler:1970hu,Beg:1974vs}. Non-perturbative long
distance hadronic corrections, were estimated by evaluating a Born
amplitude parameterized by inserting axial and vector nucleon dipole
form factors in Fig.~\ref{fig:box}, an approach introduced in
ref.~\cite{Marciano:1986pd}.

Those order $\alpha/\pi$ (with $\alpha\simeq 1/137.036$) vector and
axial-vector induced corrections, universal to all beta decays, were
estimated to be
\begin{eqnarray} \delVR = {\alpha\over 2\pi} 
\left\{   3\ln{m_Z\over m_p} + \ln{m_Z\over m_A} +2C_\text{Born} +A_g\right\},
\label{eq9}\end{eqnarray} 
The $3\ln(m_Z/m_p)$ short-distance vector current induced contribution
is free of QCD corrections, while the remaining terms, due to
axial-vector current, exhibit strong interaction effects. In
eq.~\eqref{eq9}, $m_A \sim 1.2$ GeV is a hadronic short-distance
cutoff as employed in \CMSO4, $A_g\sim -0.34$ represents its
perturbative QCD corrections and $C_\text{Born}\sim 0.86$ denotes the
Born (elastic) amplitude contribution. All three terms depend on hadronic
structure and/or perturbative QCD. Collectively, those axial-vector
induced loop contributions increase the decay rate by about
$2.9\alpha/\pi \sim 6.7\times 10^{-3}$. Although such contributions
represent a relatively minor part of the full one loop universal
radiative corrections, they carry most of the theoretical uncertainty.

A strategy for improving strong interaction effects emerged, when it
was shown \cite{Marciano:2005ec} that the perturbative QCD corrections
to beta decays and the Bjorken sum rule (BjSR)
\cite{Bjorken:1967px,Bjorken:1969mm}, the latter now known to four
loop QCD order \cite{Larin:1991zw,Larin:1991tj,Baikov:2010je}, are
identical in the chiral + isospin symmetry limits modulo small singlet
contributions that we do not consider
\cite{Larin:2013yba,Baikov:2015tea}. Therefore, one can make use of
theoretical and experimental BjSR results to define an effective
physical QCD coupling \cite{Grunberg:1980ja} that spans the
perturbative and (as more recently argued) non-perturbative loop
momentum domains and is continuous throughout the $Q^2$ transition
region (see for example
\cite{Ellis:1994py,Ellis:1995jv,Brodsky:1995tb,Deur:2005cf,Deur:2016tte,Ayala:2017uzx}). An
identical perturbative situation arises for the Gross-Llewellyn-Smith
(GLS) non-singlet sum rule \cite{Gross:1969jf}.  In fact that sum rule
is closer in structure to the $\gamma W$ box diagram and the leading
twist term in its operator product expansion.  However, accessing
relevant low $Q^2$ data in that case is less straightforward.

Employing the known BjSR or equivalent GLS non-singlet sum rule
four loop QCD corrections as input allows a
precise evaluation of the perturbative QCD corrections to the
$\gamma W$ box diagrams for loop momentum above the demarcation scale
$Q_0^2$ (see below eq.~\eqref{eq:Q0}), $Q_0^2<Q^2<\infty$, with little
uncertainty. Below $Q^2 = Q_0^2$, a non-perturbative evaluation of
hadronic loop effects is required. For that purpose, we depend
primarily on a nucleon-based form factor Born amplitude
contribution. In addition,  one of our new 
approaches, employs a
somewhat speculative analytic extension of the BjSR coupling based on
Light Front Holographic QCD (LFHQCD)
\cite{Brodsky:2003px,Brodsky:2014yha} (see also \cite{Zou:2018eam} for
a pedagogical introduction).  Our use of that non-perturbative
interpolator represents a novel application and fundamental test of
that approach. It introduces a nonperturbative $\alpha_{g_1}(Q^2)$ given
by
\begin{eqnarray} 
{\alpha_{g_1}(Q^2) \over \pi} = \exp(-Q^2/Q_0^2) \text{ for } 0<Q^2 <Q_0^2, 
\label{eq10}\end{eqnarray} 
where $g_1$ designates its dependence on the polarized structure
function $g_1(x,Q^2)$ from which it is derived (see
eq.~\eqref{eq18}). The transition scale we use,
\begin{equation}
Q_0^2=\QZeroSQRD \text{ GeV}^2,
  \label{eq:Q0}
\end{equation}
is fixed by matching non-perturbative and perturbative couplings
\cite{Herren:2017osy,Chetyrkin:2000yt} using
$\alpha_s(m_Z)= 0.1181(10)$ and the four-loop QCD code in
\cite{Chetyrkin:2000yt,Herren:2017osy}.
 The matching is quite 
smooth and leads to additional $Q^2<Q_0^2$ nonperturbative loop effects
that were neglected in 2006 \MS06\ under the assumption that they were 
included via the Born amplitude. However, as demonstrated by the DR study 
\DR\ such effects are distinct and should be separately included. 
Fortunately, they are relatively small.  Nevertheless, they are a source 
of some uncertainty and estimates of their magnitude represent the main 
difference between distinct calculations. In that regard, the DR uses
the GLS
non-singlet sum rule data at low $Q^2$ for guidance while our method
follows ideas developed from BjSR studies \MS06. Both have the same
perturbative QCD corrections modulo singlet contributions (although
the DR approach applies only three of the known four loop effects \DR)
and include similar estimates of the Born amplitude; but differ in the
low $Q^2$ evaluation of other hadronic effects. In addition to the AdS based
LFHQCD approach, we also evaluate hadronic effects using a three
resonance interpolator function fixed by boundary conditions.
Consistency of the two approaches reinforces their individual
credibility. The results are subsequently averaged to give our current
best estimate of the radiative corrections.

After presenting our updated evaluation of the RC to neutron decay, we
take this opportunity to discuss its implications for our recent
analysis of the neutron lifetime-$g_A$ connection \CMS18\ in light of
the new very precise Perkeo III \perk\ experimental result
\begin{eqnarray}  
g_A =1.2764(6) \text{ Perkeo III (2018) \protect{\perk}}, 
\label{eq11}\end{eqnarray}  
which increases the average of post 2002 experiments to 
\begin{eqnarray}  
  g_A^\text{ave} =1.2762(5) \text{ Post 2002 Experiments}. 
\label{eq12}\end{eqnarray} 
That value, taken together with the average trap neutron lifetime,
$\tau_n^\text{trap} =879.4(6)$ s is used to (conservatively) improve our previous
bound on exotic neutron decays from $<0.27\%$ to $<\bound\%$. It
actually suggests, as we later discuss, that one should probably
anticipate a future reduction in the neutron lifetime to the range 878-879
s or a decrease in the value of $g_A$.
\section{Radiative Corrections to Neutron Decay}
We begin by reviewing the electroweak radiative corrections for
neutron decay and then isolate a subset that is also universal to
super-allowed Fermi decays called $\delVR$.  The inclusive neutron
decay rate or inverse lifetime $\tau_n^{-1}$ in the SM is predicted to
be
\begin{eqnarray} {1\over \tau_n} = {G_\mu^2|\Vud|^2\over 2\pi^3} m_e^5
  \left(1+3g_A^2\right) \left( 1+\mbox{RC}\right) f,
  \label{eq13}\end{eqnarray} 
where $G_\mu = 1.1663787(6)\times 10^{-5} \text{ GeV}^{-2}$ is the
Fermi constant obtained from the muon lifetime and $g_A$ is the
axial-current coupling obtained from the neutron decay asymmetry,
$A_0=2g_A (1-g_A)/(1+3g_A^2)$. $f =1.6887(1)$ is a phase space factor
that includes the Fermi function, a relatively large roughly $+3\%$
final state enhancement due to Coulomb interactions. RC stands for
Radiative Corrections which have been taken, up until recently, to be
$+0.03886(38)$ based on a study \MS06\ in 2006. The more recent DR
approach \DR\ found $+0.03992(22)$, a significant increase outside of
the error budgets. In the case of neutron decay, RC are computed
explicitly for the vector current amplitude and $g_A$ is defined via
eq.~\eqref{eq13} so that $g_A^2$, $g_V^2$ and $f$ have the same and
factorable RC \CMSO4.  That $g_A$ as defined via eq.~\eqref{eq13} is
measured in neutron decay asymmetry studies, after correcting for
residual recoil, weak magnetism and small $\order{\alpha}$ corrections
as discussed by Wilkinson \cite{Wilkinson:1982hu} and Shann
\cite{Shann:1971fz}.  Corrections to the asymmetry reduce its
magnitude by about 1\% and correspondingly decrease $g_A$ by about
0.25\% \perk.

The purpose of this paper is to update and improve the 2006 RC
calculation approach \MS06, compare it to the recent DR result \DR, and try to
understand any remaining difference. It is predicated in part by the
DR finding that non-perturbative low $Q^2$ effects not covered by the
Born contribution are present and should be included along with a post
2006 four loop calculation of perturbative QCD corrections to the non-singlet Bj
\cite{Baikov:2010je} and GLS sum rules.

The factorized components of the lowest order RC to neutron decay are
given by 
\begin{eqnarray}
  \text{RC} = {\alpha\over 2\pi} \left[
  \overline g(E_m) + 3\ln{m_Z\over m_p} + \ln{m_Z\over m_A} 
  +2C_\text{Born}  +A_g\right], 
\label{eq14}\end{eqnarray} 
where $\overline g(E_m)$ represents long distance loop corrections as
well as bremsstrahlung effects averaged over the neutron decay
beta decay electron spectrum, and $E_m=1.292581$ MeV is the end
point electron energy specific to neutron decay.  We find its updated value is slightly
shifted to 
\begin{eqnarray} {\alpha\over 2\pi} \overline g(E_m) = 0.015035. 
  \label{eq15}\end{eqnarray} 
which reduces RC by $2\times10^{-5}$.  That contribution to the
neutron decay RC in eq.~\eqref{eq15} is specific to the neutron
spectrum and is not maintained for other beta decays, although the
function $g(E)$ used to derive it is universal to all beta decays (see
however \cite{Gorchtein:2018fxl}).  It, along with the $3\ln(m_Z/m_p)$
term in eq.~\eqref{eq9}, are independent of strong interaction effects
\cite{Sirlin:1967zza}. The rest of that RC expression represents axial
current induced effects that are dependent on strong
interactions. They provide the main focus for this paper.

\section{Axial Current Loop Contributions to RC}
The complete radiative corrections to neutron decay in
eq.~\eqref{eq9}, 
including axial-current induced and QED leading log summation effects
can be written to a good approximation as \cite{Czarnecki:2004cw}
\begin{eqnarray} 
\text{RC} = 0.03186 +1.017A_{\scriptscriptstyle
  NP}+1.08A_{\scriptscriptstyle P},
\label{eq16}\end{eqnarray} 
where the $+0.03186$ corresponds to the pure vector current
induced part of the RC including higher order effects. $A_{\scriptscriptstyle NP}$ and $A_{\scriptscriptstyle
  P}$ represent lowest order $\alpha$ long distance non-perturbative
(NP) and short distance perturbative (P) contributions to RC from
axial current effects in $\gamma W$ and $ZW$ box diagrams  (see Fig.~\ref{fig:box}).
Coefficients of the $\order{\alpha}$ contributions in eq.~\eqref{eq16}
follow from QED leading log enhancements and interference with other
parts of the vector current induced RC. The short-distance parts in
our approach correspond to loop momentum $Q^2 > Q_0^2$ (see
eq.~\eqref{eq:Q0}) while long distance parts correspond to
$Q^2<Q_0^2$.  

For the universal $\delVR$ used by Hardy and Towner 
in their analysis of super-allowed beta decays \cite{Hardy:2014qxa,Hardy:2016vhg},
one finds a corresponding approximate relationship 
\begin{eqnarray}
 \delVR = 0.01671 + 1.022A_{\scriptscriptstyle NP} +
  1.065A_{\scriptscriptstyle P}.
\label{eq17}\end{eqnarray}  
The terms in eq.~\eqref{eq16} and \eqref{eq17} were derived using the
leading log QED summation described in Appendix 1 of ref.~\CMSO4.

We subsequently employ eqs.~\eqref{eq16} and \eqref{eq17} to present
updated radiative corrections for the neutron and super-allowed beta
decays. Note, when applied to the DR $\order{\alpha}$ corrections,
eqs.~\eqref{eq16} and \eqref{eq17} give somewhat larger effects than those
reported in ref.~\DR.  However, for the most part, whenever we refer
to DR results, values cited correspond to the original literature \DR.

The short-distance axial current $\gamma W$ box diagram is the primary
source of $A_{\scriptscriptstyle P}$. It is well described using an effective
QCD coupling $\alpha_{g_1}(Q^2)$ defined for $Q^2>Q_0^2$ via
the isovector BjSR, 
\begin{eqnarray}
\int_0^1 \text{d}x\left[ g_1^p(x,Q^2) -  g_1^n(x, Q^2) \right] 
= {g_A \over 6} \left( 1 - { \alpha_{g_1}(Q^2) \over \pi} \right),
\label{eq18}\end{eqnarray} 
where $g_1$ is the polarized structure function at Bjorken $x$.  That
prescription incorporates the leading $ \order{\alpha}$ axial-current
induced amplitude from the $\gamma W$ box diagram, given by
 \begin{eqnarray} 
 \text{Box} (\gamma W)_\text{VA}= \frac{\alpha}{8\pi} \int\limits_{0}^{\infty} 
dQ^{2} \ \frac{m_{W}^{2}}{Q^{2}+m_{W}^{2}} F(Q^{2}) ,
\label{eq19}\end{eqnarray}  where the asymptotic behavior of $F(Q^2)$,
\begin{eqnarray}
F(Q^{2}) \to \frac{1}{Q^{2}}
\left( 1- \frac {\alpha_{g_1}(Q^{2})}{\pi} \right),
  \label{eq20}\end{eqnarray}
will be called the Bjorken (Bj) function,
and $\alpha_{g_1}$ is defined to be the sum of the four loop (or more if
known) QCD corrections to the BjSR \cite{Baikov:2010je},
\begin{align}
 { \alpha_{g_1} (Q^{2})\over \pi} 
& = 
a_s +(4.583 - 0.3333n_f) a_s^2 \nonumber \\
&+(41.44-7.607n_f + 0.1775n_f^2) a_s^3 \nonumber \\
&+(479.4-123.4n_f + 7.697n_f^2 - 0.1037n_f^3) a_s^4 ,
\label{eq21}
\end{align}
where $a_s = {\aMS (Q^{2}) \over \pi}$ and $n_f$ denotes the number of
(effectively massless) quark flavors.  That expression defines a
coupling $\alpha_{g_1}$ which is valid perturbatively for
$ Q^2>Q_0^2$. 
\def\skalaInt{0.9}
\begin{figure}[ht]
  \centering
\includegraphics[scale=\skalaInt]{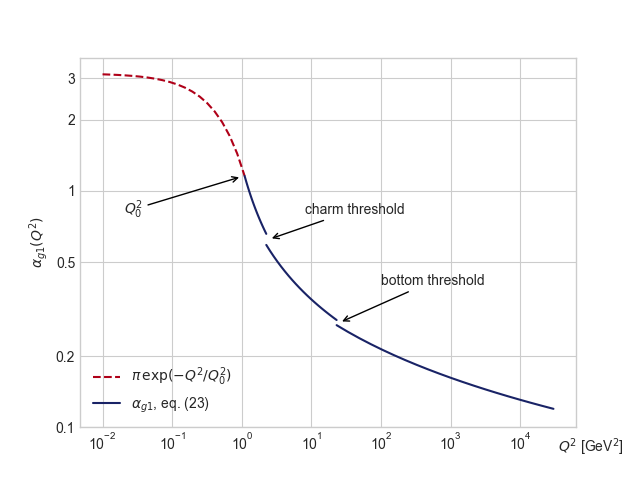} 
\caption{Effective coupling $\alpha_{g1}$ \cite{Deur:2016tte} as a
  function of $Q^2$. The nonperturbative exponential form of
  eq.~\eqref{eq22} is used for low $Q^2$ (dashed red), and the
  perturbative QCD expression \eqref{eq21} for high $Q^2$ (solid
  blue). Note the remarkably smooth matching between the two regimes.
  The discontinuities are caused by decoupling of heavy quarks
  \cite{Chetyrkin:2000yt,Herren:2017osy}.}
  \label{fig:alphaBj}
\end{figure} 

The behavior of $\alpha_{g_1}$ with $Q^2$ is shown in
Fig.~\ref{fig:alphaBj}. Discontinuities in that plot are caused by
changes in the number of active flavors in eq.~\eqref{eq21}: we change
$n_f$ when $\sqrt{Q^2}$ crosses a quark decoupling threshold. Note however
that eq.~\eqref{eq21} is derived in massless QCD. 
There are very small singlet contributions \cite{Larin:2013yba,Baikov:2015tea}
to the BjSR that enter the four loop QCD corrections.  However, at the level 
of precision we consider their effect is negligible.

In our first LFHQCD approach, for the non-perturbative domain
$Q^2<Q_0^2$, we employ the following prescription which is supported
by low $Q^2$ experimental studies of the BjSR (down to about
$Q^2 =0.2 \text{ GeV}^2$) \cite{Deur:2014vea}. We continue to use the
expression in eq.~\eqref{eq20} but with \cite{Deur:2016tte}
\begin{eqnarray} 
{\alpha_{g_1}\over \pi }= \exp (-Q^2/Q_0^2) \text{ for } Q^2<Q_0^2,
\label{eq22}\end{eqnarray} 
and
$F\left(Q_0^2=\QZeroSQRD \text{ GeV}^2\right)=\FQZeroSQRD
\text{GeV}^{-2}$, based on matching with the perturbative prediction
obtained from $\alpha_s^{\overline{\text{MS}}} (m_Z^2)$, evolved using a
five loop beta function \cite{Herren:2017osy}.  Its functional
exponential form is consistent with low $Q^2$ BjSR data and normalization
  $\alpha_{g_1}(Q^2=0)=\pi$ as suggested by AdS duality studies
  \cite{Deur:2016tte,Deur:2016cxb}. The Born (elastic) contributions
  to the hadronic corrections are computed separately using form
  factors for loop momenta $Q^2<Q_0^2$.

In 2006 \MS06, when only three loop QCD corrections to the BjSR were
known and considered, the transition $Q_0^2$ turned out to be close to
$0.7 \text{ GeV}^2$.  The use of four loop based $\alpha_{g_1}$, charm
and bottom threshold masses,  and
improved low $Q^2$ data have increased the transition $Q_0^2$ value
(see eq.~\eqref{eq:Q0}) and better establish the AdS duality
interpretation, features central to our update.

We can use the Bj function defined by eqs.~(\ref{eq20}-\ref{eq22}) to
evaluate the integral over $Q^2$ in eq.~\eqref{eq19} for the different
domains of the $\gamma W$ box diagram using
$\alpha_s(m_Z^2)= 0.1181(10)$, $m_c=1.5 \text{ GeV}$,
$m_b=4.8 \text{ GeV}$ (as decoupling thresholds of heavy quarks; see
\cite{Chetyrkin:2017lif} for an up-to-date discussion of quark masses)
and $m_t = 173.2 \text{ GeV}$ with the results:
\begin{align}
I_1 &= 0.199{\alpha\over \pi} \quad 0<Q^2<Q_0^2, \label{eq23} \\
I_2 &= \CoefITwo{\alpha\over \pi}  \quad  Q_0^2<Q^2< \infty, \label{eq24and25}
\end{align} 
where $I_i=2\times $the $\gamma W$ integrated box amplitude
contribution, as appropriate for the radiative corrections.  To those
loop effects, we must add the $ZW$ box diagram contribution
\cite{Sirlin:1974ni},
\begin{eqnarray}  
I_{\scriptscriptstyle ZW} =0.060{\alpha\over \pi},    
\label{eq26}\end{eqnarray} 
and the Born  \MS06\ integral
\begin{eqnarray}  
I_\text{Born} = \IBorn{\alpha\over \pi}   \quad 0<Q^2<Q_0^2. 
\label{eq27}\end{eqnarray} 
Contributions of QED vacuum polarization are incorporated via the
coefficients in eq.~\eqref{eq16} and \eqref{eq17}.
 
Our first estimate which follows the 2006 evaluation \MS06 but with
a four loop BjSR coupling definition and higher $Q_0^2$ value, does not
include the contribution in eq.~\eqref{eq23}.  The Born contribution
in eq.~\eqref{eq27} leads to
\begin{eqnarray}  
A_{\scriptscriptstyle NP} = \IBorn{\alpha\over \pi}   = 1.97 \times  10^{-3}, 
\label{eq28}\end{eqnarray} 
while the sum of eqs.~\eqref{eq24and25} and \eqref{eq26} gives
\begin{eqnarray}  
A_{\scriptscriptstyle P} = 2.025(21){\alpha\over \pi}   = 4.70(5) \times  10^{-3}. 
\label{eq29}\end{eqnarray} 
Plugging those values into eq.~\eqref{eq16}  and \eqref{eq17} gives
\begin{align}
\text{RC} &= \RCFirstAp, 
  \text{ First Approximation}, \label{eq30}\\
\delVR    &= \DelFirstAp,
,   \text{ First Approximation}, \label{eq30p}
\end{align}
for our updated First Approximation radiative corrections to neutron and
super-allowed nuclear beta decays.

In our next more complete AdS \cite{Deur:2016tte} motivated approach, we retain
the non-perturbative low $Q^2$ contribution from eq.~\eqref{eq23} and
find $A_{\scriptscriptstyle NP} = \newANP\times 10^{-3}$ which leads to
\begin{align}
\text{RC} & = \RCBj(32) \text{ AdS BjSR Approach}, \label{eq32}\\
\delVR    & = \DelBj(32) \text{ AdS BjSR Approach}. 
\label{eq33}\end{align} 
The two methods differ by 0.00047 with the latter about midway between
our First Approximation and the DR results \DR. The generic error
attached to eqs.~\eqref{eq32} and \eqref{eq33} as well as to later
alternative approaches, $\pm 3.2\times 10^{-4}$ corresponds to
$\pm 2.5\times 10^{-4}$ from a 10\% non-perturbative uncertainty
combined in quadrature with a $\pm 2.0\times 10^{-4}$ perturbative
error that includes QCD effects as well as uncalculated two loop
electroweak corrections and other small effects. Further reduction of that error is likely to
require a first principle's lattice calculation along with a  more complete two
loop electroweak comparison between neutron beta decay and muon decay.

The values and uncertainties given  above should be compared
with the DR results \DR,
\begin{align}
  \text{RC} & = 0.03992(22) \text{ DR result \protect\DR}, \label{eq36}\\
  \delVR &= 0.02467(22). 
\label{eq37}\end{align} 
It is interesting to contrast the AdS  (\ref{eq33}) based value,
\begin{eqnarray}
\Vud =\VudBj (18)  \text{ AdS BjSR Approach},  
\label{eq38}\end{eqnarray}  
with
\begin{eqnarray}
\Vud = 0.97370(14)  \text{ DR result \protect\DR}.
\label{eq39}\end{eqnarray} 
We note that the value of $\Vud$ in eq.~\eqref{eq38}   has moved closer
to unitarity expectations ($\sim 0.9742$). An additional shift of about
$-0.0006$ in the universal radiative corrections to super-allowed decays
or an equivalent change in another part of those studies would fully
restore unitarity. 

To examine the sensitivity of our estimate to the specific Bj function
interpolator used to integrate through the non-perturbative
$Q^2< Q_0^2$ region, we consider the resonance sum interpolator
approach introduced in 2006 \MS06 but with somewhat modified matching
conditions used to determine $F(Q^2)$ in the low momentum domain. The
new conditions allow us to better specify the non-perturbative
constraints implied by the very precise perturbative requirements.  In
that way we can match the non-perturbative and perturbative values of
$F(Q^2)$.

The underlying model of our next interpolator is large $N$ for SU($N$)
QCD, which predicts $F(Q^2)$ should correspond to an infinite sum of
vector and axial-vector resonances. As an approximation to that model,
we can use a finite sum of resonances with residues set by the
boundary and matching conditions. We can then integrate over $Q^2$
between $0$ and $Q_0^2$ i.e.~including the
non-perturbative domain in an approximation to that model. For that
purpose, we chose the sum of three resonances with residues determined
using three matching or boundary conditions. We impose conditions:
\begin{enumerate}
\item We require that in the domain $Q_0^2 \le Q^2 <\infty$ the
  three resonance
interpolator lead to the same perturbative corrections to the decay rates 
as the BjSR approach. This implies that the three resonance $F(Q^2)$ function 
satisfies the integral condition $\int _{Q_0^2}^\infty { m^2_W \over
  Q^2+m^2_W}F(Q^2)\text{d}Q^2 = 
  \ConditOne$,  four times the coefficient of $\alpha/\pi$ in $I_2$
  (cf.~eq.~\eqref{eq24and25}). We apply a condition on the integral
  rather than asymptotic  
matching in order to better reflect the effect of perturbative QCD;
\item No $1/Q^4$ terms in expansion of $F(Q^2)$ for $Q^2$ large 
or (see eq.~\eqref{eq:ABC}) $m_1^2 A + m_2^2 B +m_3^2 C
= 0$. That condition enforces chiral 
symmetry asymptotically; 
\item We employ $F(0)=A/m_1^2 +B/m_2^2 + C/m_3^2$ with $F(0)$
  arbitrary until we consider two possible ways to fix its value,
  i.e.~using either the perturbative value of $F(Q_0^2)$ or the AdS
  value of $F(0)$ as a normalization condition for the three resonance
  interpolator.
\end{enumerate}
These conditions are similar to those imposed 
in 2006 \MS06 with some improvements. Because of the larger $Q_0^2$ 
employed, the integral in condition 1 is extended down to $Q_0^2$. More 
importantly, as pointed out in the DR analysis, the condition $F(0)=0$ 
used in 2006 was not justified. Instead, we use the perturbative value of $F(Q_0^2)$
to normalize the three resonance interpolator and determine its underlying 
uncertainty.

\begin{figure}[ht]
  \centering
\includegraphics[scale=\skalaInt]{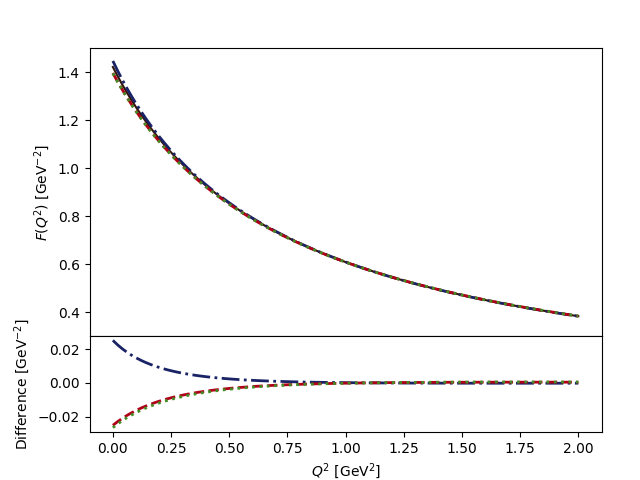} 
\caption{Interpolators as in eq.~\eqref{eq:ABC}, with $m_{1,2,3}$ as
  in eq.~\eqref{eq:massi} (solid brown line); and with $m_i$ decreased by
  5\%: $i=1$ (blue, dash-dotted), $i=2$ (red, dashed), $i=3$ (green,
  dotted). Lower panel: differences between an interpolator with a
  decreased value of $m_i$ and the interpolator with mass values given in
  eq.~\eqref{eq:massi} (the same line styles as in the upper panel).}
\label{plotF4}
\end{figure}

After solving the three coupled condition equations, one finds (for
the three resonance form of $F(Q^2)$ with given vector and axial
vector masses)
\begin{align}
 F(Q^2) &= {A \over Q^2+m_1^2} + {B\over Q^2+m_2^2} + {C \over
          Q^2+m_3^2 },    \label{eq:ABC} \\
m_1 &= \mOne \text{ GeV},\quad m_2 = \mTwo \text{ GeV}, \quad
m_3 = \mThree \text{ GeV}, \label{eq:massi}
\\
A &= \ALead(9) + \AF(3) F(0),\nonumber \\  
B & =\phantom{-}  \BLead(40)    \BF(10)F(0) ,\nonumber \\
C & =\CLead(21) + \CF(7) F(0).
\label{eq37b}\end{align}
That interpolator, integrated over $0<Q^2< Q_0^2$, leads to
\begin{equation}
  I_1(\text{three resonance}) = [\newIOnefree (9)+ \newIOneFZero (3)F(0)]{\alpha\over \pi}. 
\end{equation}
With that change in $I_1$ the radiative corrections become a function of $F(0)$,
\begin{eqnarray}
\text{RC} &=& \RCfree + \RCFZero \times 10^{-4} F(0),
\nonumber\\  
  \delVR &=& \Delfree+ \DelFZero \times 10^{-4} F(0).
\end{eqnarray}
If we match the interpolator in eq.~\eqref{eq:ABC}  with the
perturbative value $F(Q_0^2)= \FQZeroSQRD$,  we find $F(0)=\FZero$ and the 
radiative corrections which we adopt as the three resonance solution,
\begin{equation}
\left.
\begin{array}{rl}
\text{RC} =& \RCInterpol(32)  
\\  
  \delVR =& \DelInterpol(32)
\end{array}
\right\} \text{ three resonance solution},
\label{eq39b}\end{equation} 
where a small uncertainty, $\pm 4\times 10^{-5}$, is accounted for in
the $\pm 32\times 10^{-5}$ overall errors.  To test the sensitivity of
our results to the specific resonance mass scales employed, we have
redone the three resonance interpolator with each of $m_{1,2,3}$
reduced by 5\%, one $m_i$ at a time. Although the values of $A,\,B$
and $C$ are significantly modified, the different interpolators, value
of $I_1$ and radiative corrections are essentially unchanged, as
illustrated in Fig.~\ref{plotF4}. Indeed, our results are rather
insensitive to reasonable changes in the $m_i$ values.

As an alternative prescription, we evaluate the radiative corrections
resulting from the three resonance interpolator for the AdS boundary
condition $F(0) = {1\over Q_0^2}= 0.91 \gev^{-2}$ and find
\begin{eqnarray}
\text{RC} &=& \RCQZero(32) , 
\nonumber\\  
  \delVR &=& \DelQZero(32).
\label{eq39c}\end{eqnarray}
Those values are  in very good agreement with the AdS BjSR results in
eqs.~\eqref{eq32} and \eqref{eq33}. They provide
a nice consistency check on the AdS BjSR approach. We do not consider them 
as independent since both employ the same $F(0)$ boundary condition.
 
\begin{figure}[ht]
  \centering
\includegraphics[scale=\skalaInt]{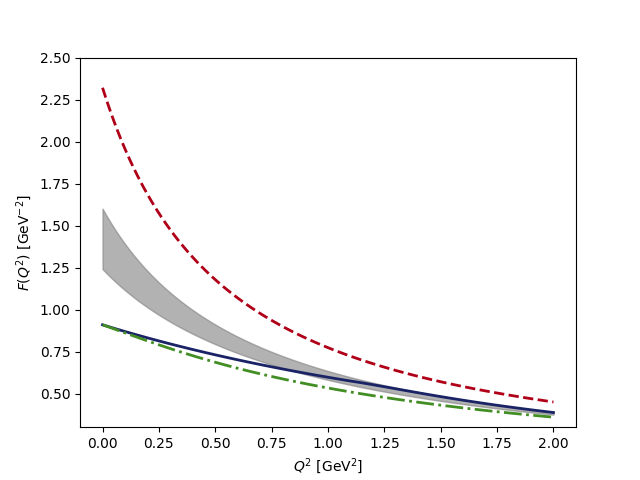} 
\caption{Examples of $F(Q^2)$ at low $Q^2$:
  $ \frac{1}{Q^{2}} \left( 1- \frac {\alpha_{g_1}(Q^{2})}{\pi}
  \right)$ (solid blue line); three resonance interpolator based on
  $F(Q^2)$ in eq.~\eqref{eq:ABC} with two choices of $F(0)$:
  $F(0)=\FZero$ (gray band) and $F(0) = 1/Q_0^2$ (dash-dotted, green);
  and two resonance interpolator from eq.~\eqref{equ45} (dashed red
  line), meant to approximate the low $Q^2$ DR findings modulo the
  Born contribution.  The area under each curve is proportional to the
  size of the radiative corrections.}
\label{plotF3}
\end{figure}

The $Q^2$ dependences of the various interpolators are illustrated in
Fig.~\ref{plotF3}.  The band surrounding the $F(0)=\FQZeroSQRD$ curve
corresponds to the uncertainty associated with the error in
$\alpha_s(m_Z^2)=0.1181(10)$. Similar bands (not shown) exist for the
other curves as well but all are small in comparison with our overall
uncertainty, $\pm 32\times 10^{-5}$ for the radiative corrections. The
good agreement between the AdS and three resonance solution for
$F(0)=0.91 \gev^{-2}$ helps validate the AdS approach. In all cases
the radiative corrections are proportional to areas under the curves.

The dashed curve in Fig.~\ref{plotF3}  corresponds to an example of 
a  two resonance interpolating function given by
\begin{equation}
         F_2(Q^2) = {1.66 \over Q^2+m_1^2} 
 - { 0.66 \over Q^2+m_2^2},   \label{equ45}
\end{equation} 
which exhibits the following features,
\begin{equation}
        F_2(0) =2.32,\quad F_2(Q_0^2)=0.732,\quad 
        F_2(2\text{ GeV}^2)=0.450. 
  \label{equ46}
\end{equation}  
It roughly represents our approximation of an effective DR
interpolator for $0<Q^2<2\gev^2$. Integrating
${\alpha\over4\pi}F_2(Q^2)$ over that domain leads to
$0.47\alpha/\pi$, in good agreement with the $0.48(7)\alpha/\pi$ found
in the DR study \DR.  Those contributions are to be compared with the
roughly $\integralToEndpt\alpha/\pi$ coming from our three resonance
interpolator when integrated over that same $Q^2$ domain. That
$\diffWithDRBelowEndpt\alpha/\pi$ difference combined with the
$\BornDifference\alpha/\pi$ Born difference $=\withBorn\alpha/\pi$ is
responsible for about a $4\times 10^{-4}$ difference between the DR
result and our three resonance interpolator finding.

The $\discrepancy \sigma$ difference between the DR and our results
may therefore be traced primarily to our use of a larger
$\alpha_s(m_Z^2)$, four loops rather than three in the QCD sum rule
corrections, different perturbative-nonperturbative matching and three
rather than two vector/axial vector poles in the interpolator.  More
specifically, to reproduce the DR low $Q^2$ contribution requires an
interpolator with an $F(0)$ central value near $2.3 \gev^{-2}$ while
our matching conditions and interpolator imply
$F(0)=\FZero \gev^{-2}$.  From that perspective it would be
interesting if a more first principles method, such as Lattice QCD,
could be employed to directly compute the value of $F(0)$.

In Table \ref{tab:1} we compare the universal and neutron specific
radiative corrections obtained from a dispersion relation approach
(line 1) with an earlier calculation from 2006 (line 2) as well as the
AdS BjSR result (line 3), three resonance interpolator (line 4), and
the average of 3 and 4 in line \lineAv.

We take the average on line \lineAv\ as representative of our study
and use it in discussing implications. Although it is somewhat smaller
than the earlier DR result \DR, they are fairly consistent.  In fact,
the agreement can be viewed as a validation of the LFHQCD and three
resonance interpolator approaches.

\begin{table}[hb]
\caption{Universal and neutron specific radiative 
corrections. \label{tab:1}}
\begin{tabular}{clll}
\hline\hline 
Line number & \hspace*{3mm} $\delVR$ &  \hspace*{3mm}   RC & Source \\
\hline 
1 & 0.02467(22)   &  0.03992(22)  & \DR\ DR Result \\
2 & 0.02361(38)   &  0.03886(38)  & \MS06 2006 Result \\
3 & \DelBj(32)   &  \RCBj(32)  & AdS BjSR Approach, eqs.~\eqref{eq32} and \eqref{eq33}  \\ 
4 & \DelInterpol(32)   &  \RCInterpol(32)  & Three Resonance Interpolator, eq.~\eqref{eq39b}\\ 
\lineAv\  & \DelAv(32)   &  \RCAv(32)  & Average of lines 3 and 4 \\
\hline \hline 
\end{tabular}
\end{table}

Although we are consistent with the DR results at about the
$\discrepancy\sigma$ level, i.e.~$\sim \absDiscrepancy \times10^{-4}$,
the remaining difference is important for interpreting CKM unitarity
and making predictions for neutron decay. In that regard, we speculate
on the basis of our analysis that the central value difference may
decrease if the DR approach is extended to include four loop QCD
corrections, low $Q^2$ corrections are parametrized using three rather
than two vector meson mass scales and perturbation theory matching is
extended below $2 \gev^2$.

We also note that the radiative corrections on lines 3, 4 and 5 of
Table \ref{tab:1} are reduced somewhat if larger values of
$\alpha_s(m_Z^2)$ are employed as input. For example, using
$\alpha_s(m_Z^2)=0.1200$, a preferred value for some experimental
inputs into the world average \cite{Tanabashi:2018oca}, leads to a
reduction by roughly $1\times 10^{-4}$ which increases our $\Vud$, as
currently extracted from super-allowed Fermi decays
\cite{Hardy:2016vhg}, from $0.97389(18)$ to $0.97394(18)$.

\section{Implications  of larger radiative corrections}
For our discussion of implications from larger radiative corrections,
we employ our averages given in line \lineAv\ of Table
\ref{tab:1}, $\delVR =\DelAv(32)$ and $\text{RC} =\RCAv(32)$. That
scenario leads to $\Vud=\VudAv(18)$ which comes closer to CKM
unitarity expectations than the DR value $\Vud=0.97370(14)$. Combined with
$|\Vus|=0.2243(9)$ from eq.~\eqref{eq4}, they correspond to roughly
\unitarityMis\ and $3.3\,\sigma$ deviations respectively. The 3.3
$\sigma$ deviation is large enough to start taking ``New Physics''
extensions of the SM seriously \cite{Belfatto:2019swo} while the
$\unitarityMis\,\sigma$ effect is more 
suggestive of missing SM 
effects. For example, nuclear physics quenching of the Born
corrections to super-allowed beta decays has been suggested as a way
of increasing $\Vud$ by about 0.00022 \cite{Seng:2018qru}. 

More specifically, if the quenching correction, as evaluated in
ref.~\cite{Seng:2018qru}, is applied to our $\Vud = \VudAv(18)$
result, it leads to $\Vud^{Q} = 0.97414(28)$ where the increased error
is due to a nuclear quenching uncertainty.  Using it together with
$\Vus = 0.2243(9)$, one finds $|\Vud^{Q}|^2 + |\Vus|^2 + |\Vub|^2 - 1 =
-0.00074(68)$, so that the first CKM row sum is consistent with unity
at close to the $1 \sigma$ level.

If instead we employ the relation \cite{Aoki:2019cca,Marciano:2004uf}
\begin{equation}
|\Vus|/|\Vud^{Q}| = 0.2313(5),\label{us2ud}
\end{equation} 
the deviation from unity is
further reduced to $-0.00028(62)$, or $-0.45 \sigma$, in excellent
agreement with CKM unitarity.  
 
As a further application, we consider the RC to neutron decay.  Using
$1+ \text{RC} = \OnePlusRCAv(32)$
 in the neutron lifetime formula
\cite{Czarnecki:2018okw}, one finds a master formula relating
$|\Vud|$, $\tau_n$ and $g_A$
\begin{eqnarray}
|\Vud|^2 \tau_n(1+3g_A^2) = 4906.4(1.7) \,\text{s}.
\label{eq40b}\end{eqnarray}  
Employing $\tau_n^\text{trap}=879.4(6)$ s and post 2002 average
$g_A$=1.2762(5) leads to
\begin{eqnarray}
  \Vud = 0.9736(5) .
\label{eq41}\end{eqnarray} 
The uncertainty in $\Vud$ from neutron decay measurements is starting
to become competitive in accuracy with super-allowed beta decay
determinations. In addition, its central value may also be indicating
a deviation from unitarity. A central value shift to unitarity and
$\Vud \sim 0.9742$ would require a reduction in either $\tau_n$ or
$g_A$. Given the recent precision of Perkeo III, we consider $g_A$
fixed at the new post 2002 average 1.2762(5) which then suggests a
$\tau_n < 879$~s.

An alternate interpretation of the apparent violation of CKM unitarity in
eq.~\eqref{eq8}  resulting from larger universal radiative corrections,
consistent with $|\Vus|/|\Vud|=0.2313$ from $K_{\mu2}/\pi_{\mu2}$, suggests
the solution $\Vud=0.9735$ and $\Vus=0.2252$ which requires ``new
physics.''
One possible explanation 
could be the existence of a $0.1\%$ 
increase in the muon decay rate from ``new
physics'' which shifts $G_\mu$ to a 
value larger than the real $G_F$.  Alternatively, it could stem from an 
opposite sign effect in nuclear beta decay.
That solution agrees with the current central value in eq.~\eqref{eq41}.
Of course, such a scenario would be very exciting. It will also be well 
tested by the next generation of precise $\tau_n$ and $g_A$ measurements.

Recently, we discussed  \CMS18\ a resolution of the neutron lifetime problem
(the beam $\tau_n^\text{beam}=888.0(2.0)$ s and trap
$\tau_n^\text{trap}=879.4(0.6)$ s lifetime discrepancy) based on a
precise connection between $\tau_n$ and $g_A$, the axial coupling
measured in neutron decay asymmetries. We note that a shift in the
universal beta decay radiative corrections alone makes a negligible
change in the relationship
\begin{equation}
 \tau_n(1 + 3g_A^2) = 5172.0(1.1) \text{ s}, 
\label{eq42}\end{equation}
used for that study due to a cancelation of uncertainties and common
shifts between super-allowed and neutron beta decay rates. Similarly, a
change in $G_F$ will not change eq.~\eqref{eq42}.  However, a shift in
the nuclear theory corrections as suggested in \cite{Seng:2018qru}
will modify it. For example, a shift in $\Vud$ by $+0.0002$ by further
quenching of the Born contribution would lower the 5172.0 s in
eq.~\eqref{eq42} to 5169.9 s. More important is the recent increase in
the post 2002 $g_A^\text{average}$ from 1.2755(11) to 1.2762(5) with
the addition of the Perkeo III result \perk\ in eq.~\eqref{eq11}. That
shift reduces the predicted neutron lifetime from 879.5(1.3) s to
\begin{equation}
\tau_n =878.7(0.6) \text{ s } (\text{prediction
  based on } g_A=1.2762 (5)).
\label{eq43}\end{equation} 
That prediction is further reduced if $\Vud$ were to increase to
respect CKM unitarity. Indeed, one would expect $\tau_n$ closer to 878
s. 

We conclude by noting that the new post 2002 $g_A$ average in
eq.~\eqref{eq12} can be used in the analysis of \CMS18 to reduce the
bound on exotic neutron decays (such as $n\to \text{dark particles}$
\cite{Fornal:2018eol,Berezhiani:2018udo}) from $<0.27\%$ to
\begin{equation}
  \BR(\text{exotic neutron decays}) < \bound\% \text{ (95\% one-sided CL)}, 
\label{eq44}\end{equation}
where we have not allowed 
for negative exotic branching ratios in the statistical distribution.
That bound leaves little chance for a  1\% dark particle decay as the solution to the
neutron lifetime problem (unless one modifies the neutron asymmetry
with new physics, e.g.~ref.~\cite{Ivanov:2018vit}).

We have presented an updated analysis of the radiative corrections to 
neutron and super-allowed nuclear beta decays. It extends the BjSR 
function into the non-perturbative low loop momentum region, incorporating
four loop QCD effects as well as LFHQCD ideas and their confirmation by low 
energy experimental data.  The value obtained was averaged with  a slightly 
larger three resonance result.  On the basis of our considerations, we advocate
the universal value $\delVR =+\DelAvPerc(32)\%$ as a competitive result
about midway between earlier estimates \MS06 and the recent dispersion relation 
result \DR.   Further study of the remaining small difference is 
warranted.  Tests of both approaches will result from the next generation
of neutron lifetime and $g_A$ asymmetry measurements that aim for $10^{-4}$ 
sensitivity.  Lattice calculations of $F(Q^2)$ may be possible \cite{Seng:2019plg}.
Will CKM unitarity be violated and
``New Physics'' uncovered? Time will tell.

Acknowledgement:\\ We thank Konstantin Chetyrkin for helpful remarks
on the manuscript. W.~J.~M.~thanks C.~Y.~Seng, M.~Gorchtein and
M.~J.~Ramsey-Musolf for discussions. The work of A.~C.~was supported
by the Natural Sciences and Engineering Research Council of
Canada. The work of W.~J.~M.~was supported by the U.S. Department of
Energy under grant DE-SC0012704. The work of A.~S.~was supported in
part by the National Science Foundation under Grant PHY-1620039.

\def\disclaim{0}
\ifnum \disclaim=1
Notice: This manuscript has been co-authored by employees of
Brookhaven Science Associates, LLC under Contract No. DE-SC0012704
with the U.S. Department of Energy. The publisher by accepting the
manuscript for publication acknowledges that the United States
Government retains a non-exclusive, paid-up, irrevocable, world-wide
license to publish or reproduce the published form of this manuscript,
or allow others to do so, for United States Government purposes.  This
preprint is intended for publication in a journal or
proceedings. Since changes may be made before publication, it may not be cited or reproduced without the author's permission.
\subsection*{DISCLAIMER}

This report was prepared as an account of work sponsored by an agency
of the United States Government. Neither the United States Government
nor any agency thereof, nor any of their employees, nor any of their
contractors, subcontractors, or their employees, makes any warranty,
express or implied, or assumes any legal liability or responsibility
for the accuracy, completeness, or any third party's use or the
results of such use of any information, apparatus, product, or process
disclosed, or represents that its use would not infringe privately
owned rights. Reference herein to any specific commercial product,
process, or service by trade name, trademark, manufacturer, or
otherwise, does not necessarily constitute or imply its endorsement,
recommendation, or favoring by the United States Government or any
agency thereof or its contractors or subcontractors. The views and
opinions of authors expressed herein do not necessarily state or
reflect those of the United States Government or any agency thereof.
\fi

\end{document}